\documentclass{jfm}
\usepackage{graphicx}
\usepackage{natbib}
\usepackage{amsbsy}
\usepackage{amsmath}
\usepackage{amssymb}
\usepackage{bm}
\usepackage{latexsym}
\usepackage{color}

\title[Large-scale mean patterns in turbulent convection]{Large-scale mean patterns in turbulent convection}

\author[Mohammad S. Emran and J\"org Schumacher]{Mohammad S. Emran and   J\"org Schumacher}
 \affiliation{Institut f\"ur Thermo- und Fluiddynamik, Postfach 100565, Technische Universit\"at Ilmenau, D-98684 Ilmenau, Germany}
\date{\today}
\begin{document}
\maketitle

\begin{abstract}
Large-scale patterns, which are well-known from the spiral defect chaos regime of thermal convection at 
Rayleigh numbers $Ra < 10^4$, continue to exist in three-dimensional numerical simulations of turbulent Rayleigh-B\'{e}nard convection 
in extended cylindrical cells with an aspect ratio $\Gamma=50$ and $Ra>10^5$. They are uncovered when the turbulent fields are averaged in time and turbulent 
fluctuations are thus removed. We apply the Boussinesq closure to estimate turbulent viscosities 
and diffusivities, respectively. The resulting turbulent Rayleigh number $Ra_{\ast}$, that describes the convection of the mean
patterns, is indeed in the spiral defect chaos range. The turbulent Prandtl numbers  are smaller than one with 
$0.2\le Pr_{\ast}\le 0.4$ for Prandtl numbers $0.7 \le Pr\le 10$. Finally, we demonstrate that these mean flow patterns are robust to an additional 
finite-amplitude side wall-forcing when the level of turbulent fluctuations in the flow is sufficiently high. 
\end{abstract}


\section{\label{sec:level1} Introduction}
The formation of regular patterns close to the onset of a hydrodynamic instability in spatially extended flows is well documented for generic cases. The most 
prominent examples are convection rolls in Rayleigh-B\'{e}nard flow (\cite{Busse1978}, \cite{Bodenschatz2000}) heated from below and cooled from above, 
Taylor vortices in Taylor-Couette flow (\cite{Andereck1986}) between two rotating concentric cylinders and inclined turbulent stripe patterns in plane-shear flows driven by a 
pressure gradient or a wall movement (\cite{Barkley2005}, \cite{Duguet2013}). Specifically linearly unstable systems, such as Rayleigh-B\'{e}nard convection, with 
a sharp transition threshold to the convective flow state allow then for a perturbative expansion about the first unstable mode at onset. The expansion leads to an amplitude 
equation which is simpler than the original fluid equations and describes the formation of simple patterns as a function of the system parameters (\cite{Cross1993}, \cite{Hoyle2006}). 
The derivation of nonlinear phase diffusion equations allows to model increasingly complex patterns, such as spirals or defects (\cite{Hoyle2006}), which have also been detected 
in experiments, e.g., by \cite{Croquette1989} and \cite{Morris1991}. Defects are imperfections in the patterns such as dislocations (\cite{Cross2009}). In large-aspect ratio cells, this results
in a state of slowly evolving spirals and defects which is known as spiral defect chaos (SDC). In such SDC regimes, all symmetries of the governing equations  have been 
spontaneously broken (\cite{Busse2003}).  For example, the azimuthal symmetry of the roll patterns is broken in an extended cylindrical cell. When the temperature difference across 
the fluid layer is further increased, the fluid motion crosses over from the weakly nonlinear to the turbulence regime. 

The dimensionless  Rayleigh number $Ra=g\alpha\Delta T H^3/(\nu \kappa)$ describes the thermal driving. It contains the acceleration due to gravity, $g$, the thermal expansion coefficient at 
constant pressure, $\alpha$, the outer sustained temperature difference across the layer, $\Delta T$, and the height of the layer (or convection cell), $H$. The kinematic viscosity 
of the fluid $\nu$ and the thermal diffusivity $\kappa$ of the temperature field form the second important dimensionless parameter which relates dissipation in the working fluid to thermal diffusion, 
the Prandtl number $Pr=\nu/\kappa$. Turbulence is characterized by an irregular, stochastic and three-dimensional fluid motion. Does this however imply that the 
patterns for velocity and temperature which are documented in the SDC regime at lower Rayleigh number disappear? If not, how can these patterns be extracted? How robust are they 
with respect to variations of the Prandtl number? And finally, how robust are they with respect to an additional side wall-forcing which is added
to the momentum equation?  These are the questions which we want to address in the present work.  

Our investigation is based on three-dimensional direct numerical simulations (DNS) in very large aspect ratio cells which are comparable to 
laboratory experiments in pattern formation (\cite{Bodenschatz2000}).  We conducted a series of DNS of the Boussinesq equations in cylindrical 
cells with an aspect ratio of $\Gamma=D/H=50$ with $D$ being the diameter of the cell. Our investigation extends previous numerical studies of turbulent 
Rayleigh-B\'{e}nard convection by \cite{Hartlep2005}, \cite{Hardenberg2008} and \cite{Bailon2010} to very large aspect ratios $\Gamma$ (see also \cite{Chilla2012}).  We show that 
patterns very similar to SDC continue to exist into the soft turbulence regime, up to Rayleigh numbers $Ra=500\,000$ which were accessible here. 
The patterns are covered by an increasing amplitude of velocity fluctuations and become visible only after a time averaging over a sequence of flow snapshots. We then 
determine turbulent viscosities, $\nu_{\ast}$, and diffusivities, $\kappa_{\ast}$, for the time-averaged roll patterns and define turbulent Rayleigh and Prandtl numbers 
for the mean flow patterns. That means we replace
\begin{equation}
Ra\to Ra_{\ast}=\frac{g\alpha\Delta T H^3}{\nu_{\ast} \kappa_{\ast}}\ll Ra\;\;\;\;
\mbox{and}\;\;\;\;Pr\to Pr_{\ast}=\frac{\nu_{\ast}}{ \kappa_{\ast}}\sim {\cal O}(1)\,.
\end{equation}
These turbulent Rayleigh numbers are significantly smaller than the original ones and fall consistently back into a range that corresponds with the original spiral defect chaos 
regime. The turbulent Prandtl numbers decrease as well and remain smaller than one. The value of $Pr_{\ast}$ depends weakly on $Pr$. It increases with increasing 
$Pr$.    

\section{Numerical model}
We perform DNS of the three-dimensional Boussinesq equations which are given by 
\begin{eqnarray}
\label{nseq}
\frac{\partial u_i}{\partial t} +u_j\frac{\partial u_i}{\partial x_j}
&=&-\frac{1}{\rho_0}\frac{\partial p}{\partial x_i} +\nu \frac{\partial^2 u_i}{\partial x_j^2}+\alpha g (T-T_0) \delta_{3i} + f_i\,,\\
\label{ceq}
\frac{\partial u_j}{\partial x_j}&=&0\,,\\
\frac{\partial T}{\partial t}+u_j \frac{\partial T}{\partial x_j}
&=&\kappa \frac{\partial^2 T}{\partial x_j^2}\,,
\label{pseq}
\end{eqnarray}
where $p(x_k,t)$ is the pressure, $u_i(x_k,t)$ the velocity field, $\rho_0$ the constant mass density, and $T(x_k,t)$ the temperature field. The temperature $T_0$ is a reference 
temperature. Summation is applied over index $j$ in Eqns. (\ref{nseq}) -- (\ref{pseq}), where $i,j,k=1,2,3$. The last term on the right hand side of Eq. (\ref{nseq}), which is denoted by $f_i(x_k,t)$, stands 
for an additional volume forcing. It is applied close to the side walls of the convection cell and designed such that it enforces the azimuthal symmetry in the vicinity of the side walls. This 
additional forcing is applied for two runs only which are discussed in section 3.3. 

The velocity field has a no-slip boundary condition on all walls. The temperature boundary condition is isothermal at the top and bottom plates and adiabatic at the side wall. The problem is 
formulated in cylindrical coordinates $(r, \phi, z)$ and solved by a second-order finite difference scheme (\cite{Verzicco2003}).  In Table \ref{Tab1}, we list the simulation parameters and 
grid resolutions for each DNS run. The azimuthal spacing ($\Delta_{\phi}$) is uniform, the radial ($\Delta_r$) and axial ($\Delta_z$) grid sizes are nonuniform. The radial mesh gets finer towards 
the side wall. It is obtained by a geometric scaling relation which clusters the grid points less than Tchebychev collocation points when the side wall is approached, $r\to D/2$.
This also means that azimuthal grid spacing $r\Delta_{\phi}$ grows for growing $r$. The grid resolution in the cylindrical cell with very large aspect ratio puts a challenge to the simulations, 
in particular in terms of the azimuthal resolution.  Following \cite{Groetzbach1983}, we tested our DNS grid by calculating the global maximum of the geometric mean, $\tilde{\Delta}=\max(\sqrt[3]{r\Delta_{\phi}
\Delta_r\Delta_z})$. The criterion states that $\tilde{\Delta}\le \pi\eta_K$, where $\eta_K$ is the Kolmogorov  dissipation length (see also \cite{Emran2008} and \cite{Shishkina2010}). 
The ratio $\tilde{\Delta}/\eta_K$ is 2.7 for run 2 in Table \ref{Tab1} and 1.7 for comparison run 2a at a higher resolution.  Turbulent heat and momentum transfer are compared in Tab. \ref{Tab1}. 

All runs start with the diffusive equilibrium state which is perturbed randomly. Length scales are normalized in 
units of $H$,  velocities in units of the free-fall velocity $U_f=\sqrt{g\alpha\Delta T H}$, time in units of the free-fall time $T_f=H/U_f$ and temperatures in units of $\Delta T$. The turbulent heat and
momentum transport are measured by the Nusselt and Reynolds numbers, respectively. They are given by 
\begin{equation}
Nu=1+\sqrt{Ra Pr} \langle \tilde{u}_z \tilde{T}\rangle_{V,t}\,,\;\;\;\;\;\;\;Re= \sqrt{\frac{Ra}{Pr} \langle \tilde{u}^2_i\rangle_{V,t}}\,.
\end{equation} 
The notion $\langle\cdot\rangle_{V,t}$ stands for an ensemble average taken as a volume-time average in DNS case. From here on, we 
will omit the tilde for dimensionless quantities. 
\begin{table}
\begin{center}
\begin{tabular}{ccrrrrrrrr}
\hline
$Run$ & $N_{\phi}\times N_r\times N_z$ & $Ra$ & $\Gamma$ & $Pr$ &  $Nu$ & $Re$ & $N_{BL}$ & $Ra_{\ast}$ & $Pr_{\ast}$  \\
\hline
1 & $601\times401\times 97$   & $5\,000$  & 50 & 0.7  & 1.84  & 14 & 30 & -- & --\\
2 & $601\times401\times97$    & $500\,000$  & 50 & 0.7  &  7.72 & 204 & 10 & 4500 & 0.21\\
2a & $1201\times601\times141$    & $500\,000$  & 50 & 0.7  &  7.25 & 209 & 16 & 4800 & 0.21\\
3 & $601\times401\times97$    & $500\,000$  & 50 & 3   & 8.38  & 62  & 10 & 7300 &  0.38\\
4 & $601\times401\times97$    & $5\,000$  & 50 & 10   &  2.00 & 1 & 29 & -- & --\\
5 & $601\times401\times97$    & $500\,000$  & 50 & 10   &  8.71 & 22  & 9 & 39000 & 0.40 \\
\hline
\end{tabular}  
\end{center}
\caption{Parameters of the different simulations. The column $N_{BL}$ displays the number of grid points
inside the thermal boundary layer. The last two columns display the turbulent Rayleigh and Prandtl 
numbers, respectively, analyzed for selected runs. Run 2a is conducted at a higher grid resolution
as run 2.}
\label{Tab1}
\end{table}
\begin{figure}
\begin{center}
\includegraphics[scale=0.15]{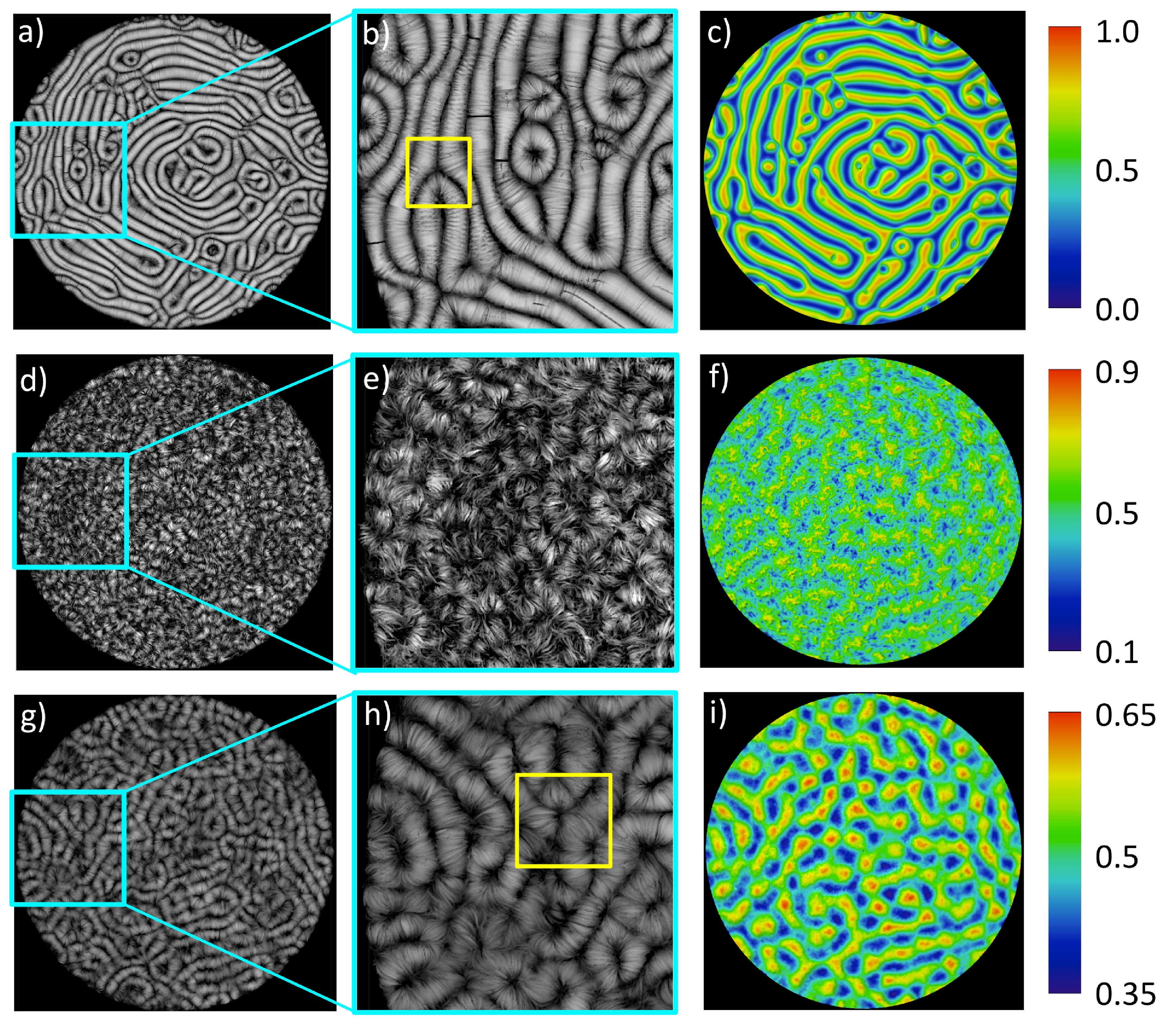}
\caption{Streamlines of the velocity field (view from the top) and contours of the temperature field in a Rayleigh-B\'{e}nard convection cell. (a) Instantaneous
velocity field pattern at $Ra=5\,000$ and (b) magnification, both taken from run 1 in table \ref{Tab1}. (c) Corresponding temperature field in mid plane. (d,e) Instantaneous 
streamline plot and its magnification at $Ra=500\,000$. (f) Corresponding temperature field in mid plane. (g,h) Streamline plot of the time-averaged velocity field at 
$Ra=500\,000$ and magnification. (i) Corresponding time-averaged temperature field in mid plane. The time average in (g)--(i) is taken over $\tau=200 T_f$. Panels (d)--(i) are for run 2 
from table \ref{Tab1}. All data are for $Pr=0.7$. The yellow boxes in panels (b) and (h) highlight defects in the patterns.}  
\label{fig1}
\end{center}
\end{figure}

\section{Results}
\subsection{Observations}
Figure \ref{fig1} shows a sequence of three-dimensional streamline plots viewed from the top for $Pr=0.7$ and $\Gamma=50$.  Panel (a) and a magnification in (b) are 
for $Ra=5\,000$. For this Rayleigh number value almost no difference was found between an instantaneous snapshot and the time average which is taken over 100 $T_f$ and not shown in the figure.
The corresponding temperature pattern is displayed in panel (c). Panel (d) of the same figure and its magnification (e) display an instantaneous streamline 
plot at $Ra=500\,000$. Both figures reflect the large amplitude of turbulent fluctuations. The fluctuating nature of the temperature field is also obvious in Fig.
\ref{fig1}(f). The snapshots appear at a first glance almost featureless. The bottom panels (g)--(i) show the time averages, which are obtained for a duration of 
$200 T_f$, and its magnification. The temperature plots (f) and (i) recapture patterns which have been discussed in \cite{Hartlep2005} for similar Rayleigh and Prandtl numbers
in rectangular slabs with $\Gamma=10$. This holds particularly in the center of the convection cell. The magnified view of panel (g) in (h)  confirms the well-known result that the mean flow rolls 
end perpendicular to the side wall which underlines that the grid resolution is sufficient. 

The time-averaged plots (g)--(i) recapture now patterns that are similar to SDC, i.e., to those which are observed in panels (a)--(c) of the same figure for the Rayleigh 
number that is by two orders of magnitude smaller.  A time average taken over $\tau$ has to be long enough such that the turbulent fluctuations in the velocity field are suppressed 
$(\tau\gg T_f)$. However, if the averaging procedure proceeds over a very long time interval then these patterns will be washed out for all Rayleigh numbers discussed here.
\begin{figure}
\begin{center}
\includegraphics[scale=0.13]{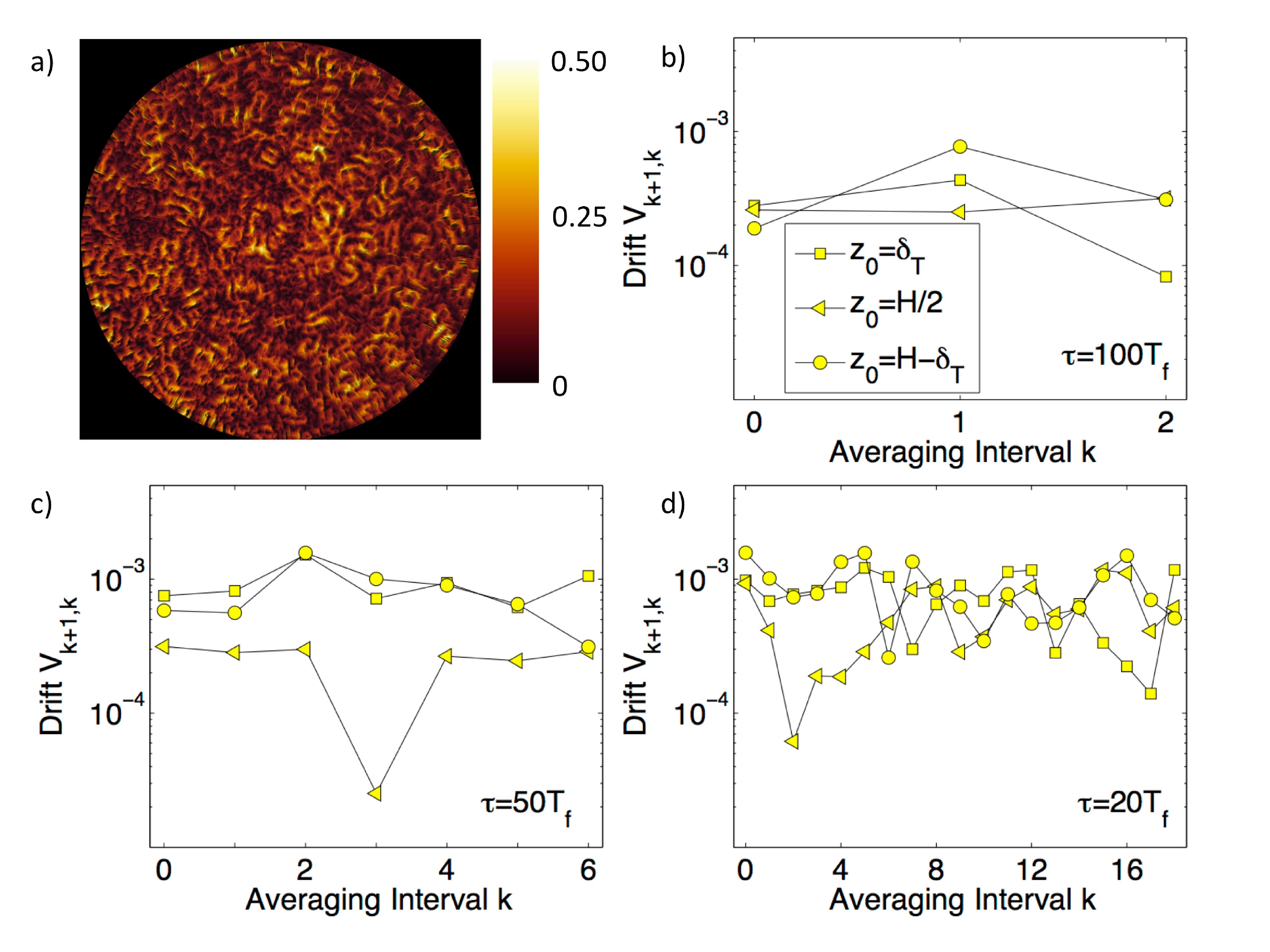}
\caption{(Color online) Determination of the drift between two successive mean flow patterns in order to quantify the slow variation. (a): Magnitude of the difference between two 
successive mean flow patterns taken at $z_0=H/8$ with $\tau=50 T_f$. (b--d): $V_{k+1,k}(z_0)$ versus averaging interval $k$ taken at three different $z_0$ which are indicated 
in the legend and the same for all three panels. Data are for run 2 in table \ref{Tab1}. (b) $\tau=100 T_f$, (c) $\tau=50 T_f$, (d) $\tau=20 T_f$.}
\label{fig1b}
\end{center}
\end{figure}
We can decompose the velocity and temperature fields into a time-averaged field and remaining 
turbulent fluctuations as
\begin{equation}
u_i(x_j,t)=\langle u_i(x_j)\rangle_{t}+u_i^{\prime}(x_j,t)\;\;\;\;\;\mbox{and}\;\;\;\;\;
T(x_j,t)=\langle T(x_j)\rangle_{t}+T^{\prime}(x_j,t)\,,
\label{decompo}
\end{equation}  
where $\langle\cdot\rangle_{t}$ denotes a time average. In Fig. \ref{fig1b}, we analyze the slow drift of the large-scale flow pattern. The total integration time 
interval, $T=M\tau$, is divided into $M$ equidistant subintervals ${\cal I}_{k}$ with $k=0, M-1$. These averaging intervals are taken from $k \tau$ to $(k+1)\tau$ with $\tau\gg T_f$.  Panel (a) 
of the figure shows the magnitude of the difference between two successive mean flow patterns taken at $z_0=H/8$. We see that pointwise differences get as high as 
$0.5 U_f$ in this example. Panels (b)--(d) of Fig. \ref{fig1b} show a measure for the drift of the mean flow patterns which is defined as
\begin{eqnarray}
V_{k+1,k}(z_0)=\Big|\langle u_i(z_0)\rangle_{A,t\in {\cal I}_{k+1}}-\langle u_i(z_0)\rangle_{A,t\in {\cal I}_{k}}\Big|\,.
\end{eqnarray}  
The notation $\langle\cdot\rangle_{A,t}$ stands for a plane-time average. Run 2 is advanced for $400 T_f$ and 
this time interval is split into fractions of $\tau= 20, 50$ and $100 T_f$, respectively. While for $\tau=100 T_f$ the drift velocities in all three planes are of the same size, the data for the two 
smaller $\tau$ imply that the drift in the center plane is in parts slightly slower. For averaging times $\tau\lesssim 20 T_f$ we reach the range of typical turnover times of a Lagrangian tracer within a 
large-scale circulation roll (\cite{Emran2010}). Therefore, we do not consider smaller time intervals $\tau$. In all three plots, we detect nearly the same magnitude of the drift velocity. This 
allows us to derive a time scale of the processes which is $H/V_{k+1,k}\gtrsim 10^3 T_f$, a large time scale which is not accessible in this study. This time scale 
is comparable to that of a slow spanwise drift of streaky structures in plane Poiseuille flow which has been reported very recently by \cite{Kreilos2014}. This estimate
is also consistent with the one for a time scale of horizontal motion, $T_h$, that should vary as $T_h=\Gamma^2 T_f$.
Furthermore, we observe that the drift velocities for planes at $z=\delta_T,\, H/2$ and $H-\delta_T$ are of same order of magnitude. This suggests that the 
mean flow roll pattern drifts slowly as a whole.  

\begin{figure}
\begin{center}
\includegraphics[scale=0.14]{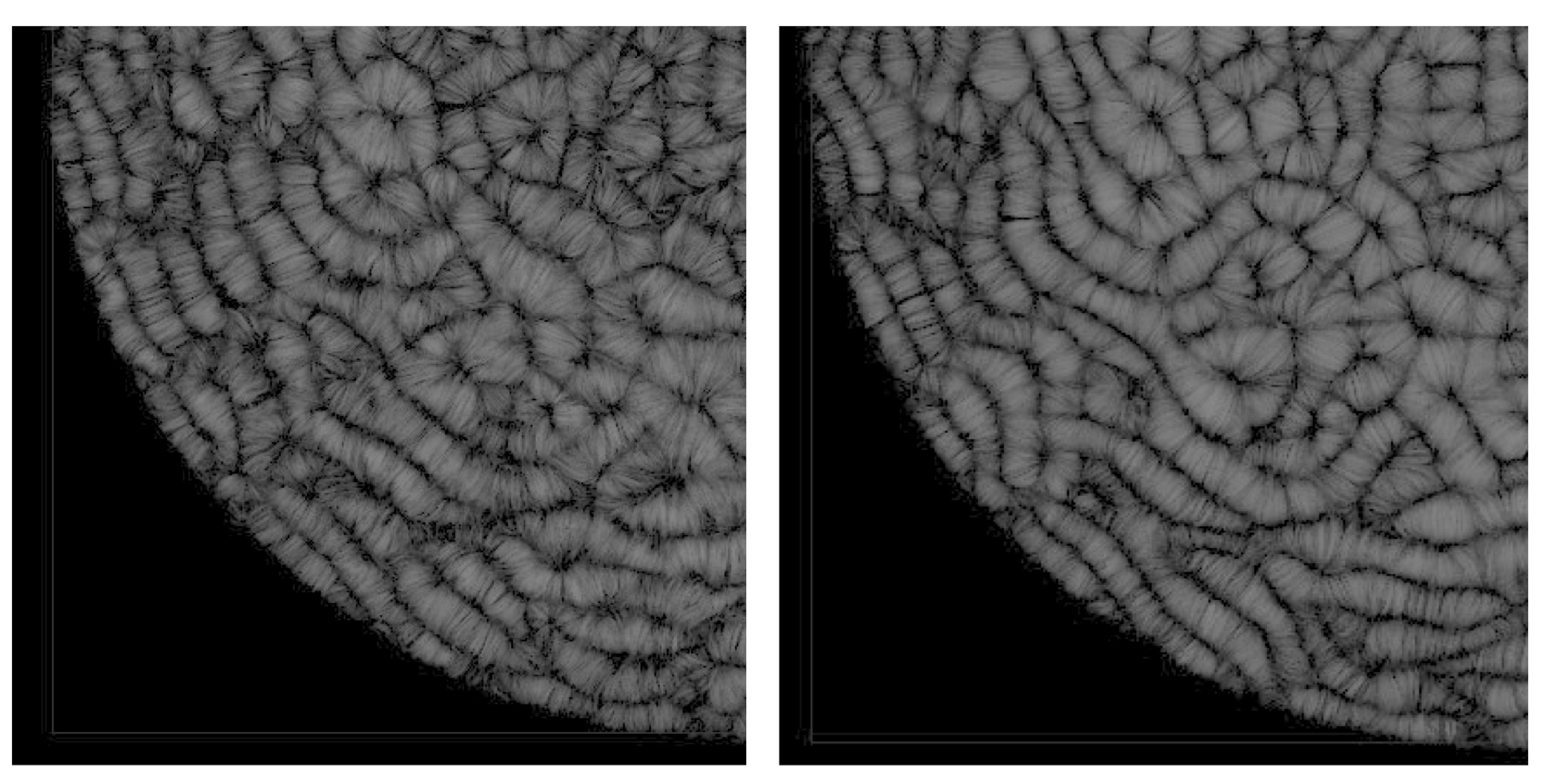}
\caption{Streamline plots of the velocity field in the Rayleigh-B\'{e}nard convection cell at $Ra=500\,000$ and $Pr=10$. A view from the top onto a quarter 
of the cell is displayed. Left: instantaneous streamline snapshot. Right:  time-averaged streamline plot obtained for averaging over $\tau=50 T_f$. Data are from run
5 in table \ref{Tab1}.}  
\label{fig2}
\end{center}
\end{figure}

Figure \ref{fig2} repeats the analysis at $Pr=10$ and $Ra=500\,000$. Now, the streamlines of the  
snapshot appear much less disordered than for $Pr=0.7$. Consequently, the difference to the mean flow pattern is much smaller. One reason could be that the thermal diffusion is less compared to the momentum diffusion when the Prandtl number grows for a fixed Rayleigh number. This results in thermal plumes which have thinner stems and disperse less rapidly with respect to time. Thus the 
stirring of the fluid by plumes is less efficient. The result is in line with the decrease of the Reynolds number for growing Prandtl number as shown in Tab. \ref{Tab1}. Our
finding is also supported by \cite{Silano2010} who have observed decreasing peak velocities for increasing $Pr$. 

\begin{figure}
\begin{center}
\includegraphics[scale=0.6]{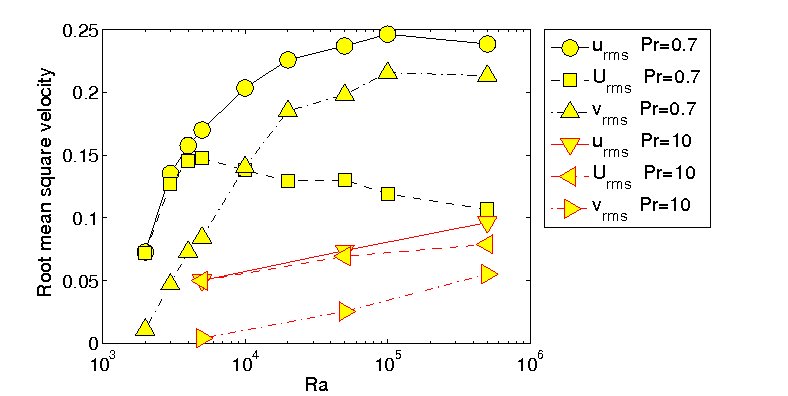}
\caption{(Color online) Root mean square values of the total velocity, the time-averaged velocity and the remaining turbulent fluctuations as a function of Rayleigh number $Ra$ (see Eqns. (\ref{rms})).
Additional data points beside those listed in Tab. \ref{Tab1} are given at $Ra=2\,000, 3\,000, 4\,000, 10\,000, 20\,000, 50\,000$ and $100\,000$ for the series at $Pr=0.7$ and 
$Ra=50\,000$ for $Pr=10$.}  
\label{fig3}
\end{center}
\end{figure}

In Fig. \ref{fig3}, we summarize the results of the Reynolds-decomposed velocity field (see the decomposition in Eq. (\ref{decompo})). In detail, we define 
\begin{equation}
{u}_{rms}=\sqrt{\langle u_i^2\rangle_{V,t}}\,,\;\;\; {U}_{rms}=\sqrt{\langle \langle u_i\rangle_{t}^2\rangle_{V}}\,,\;\;\;{v}_{rms}=\sqrt{\langle u_i^{\prime\,2}\rangle_{V,t}}\,.
\label{rms}
\end{equation}  
We include further runs at the same resolution which are not listed in Table \ref{Tab1}, but in the caption. The smallest Rayleigh number was $Ra=2\,000$ for $Pr=0.7$ which
is slightly larger than the linear instability threshold, $Ra_c=1\,708$. When expressed as a distance to the linear instability threshold this gives $\varepsilon=(Ra-Ra_c)/Ra_c=0.17$.  
In this case, velocity fluctuations are practically absent, the flow pattern is almost steady consisting of several subdomains with stripe textures.  With increasing Rayleigh 
number fluctuations of all three parts of the velocity field  (see Eqns. (\ref{rms})) grow up to $Ra\approx 5\,000$ which corresponds to $\varepsilon=(Ra-Ra_c)/Ra_c=1.93$. At about this 
Rayleigh number, $U_{rms}$ reaches a local maximum and starts to decrease with increasing Rayleigh number. At $Ra\approx 10\,000$, the 
turbulent fluctuations $v_{rms}$ exceed $U_{rms}$. At $Ra\sim 100\,000$, $u_{rms}$ and $v_{rms}$ reach a local maximum and level off. For this Rayleigh number, the 
flow is already turbulent, the fluctuations $v_{rms}$ are by a factor of two larger than $U_{rms}$.    
We also show three data sets for the case of $Pr=10$. The magnitudes of all three parts are significantly reduced which confirms our observation from Fig. \ref{fig2}. 
Up to the accessible $Ra=500\,000$ all three terms continue to grow suggesting that the maxima are shifted to higher $Ra$. 
  
\subsection{Estimate of turbulent viscosity and diffusivity}
The next step is to estimate the turbulent viscosities and diffusivities in the bulk of the cell and to evaluate the resulting turbulent Rayleigh and Prandtl numbers. 
We start with the Boussinesq ansatz for the closure which connects turbulent fluxes (or stresses) with the mean gradients (see e.g. \cite{Wilcox2006,Shams2014}) and states that
\begin{equation}
\overline{u_i^{\prime} u_j^{\prime}} =-\nu^{ijkl}_{\ast} \frac{\partial \overline{u}_l}{\partial x_k}\,,\;\;\;\;\;\;\;    
\overline{u_i^{\prime} T^{\prime}} =-\kappa^{ij}_{\ast} \frac{\partial \overline{T}}{\partial x_j}\,, 
\label{Boussinesqanasatz}
\end{equation}  
where bars denote an appropriate space-time average. Our following estimate will aim at obtaining numbers $\nu_{\ast}$
and $\kappa_{\ast}$ rather than exploring the full tensorial structure of the turbulent viscosities and diffusivities. This would go
beyond the scope of this work. We will restrict the analysis to the dominant contributions only.

\begin{figure}
\begin{center}
\includegraphics[scale=0.11]{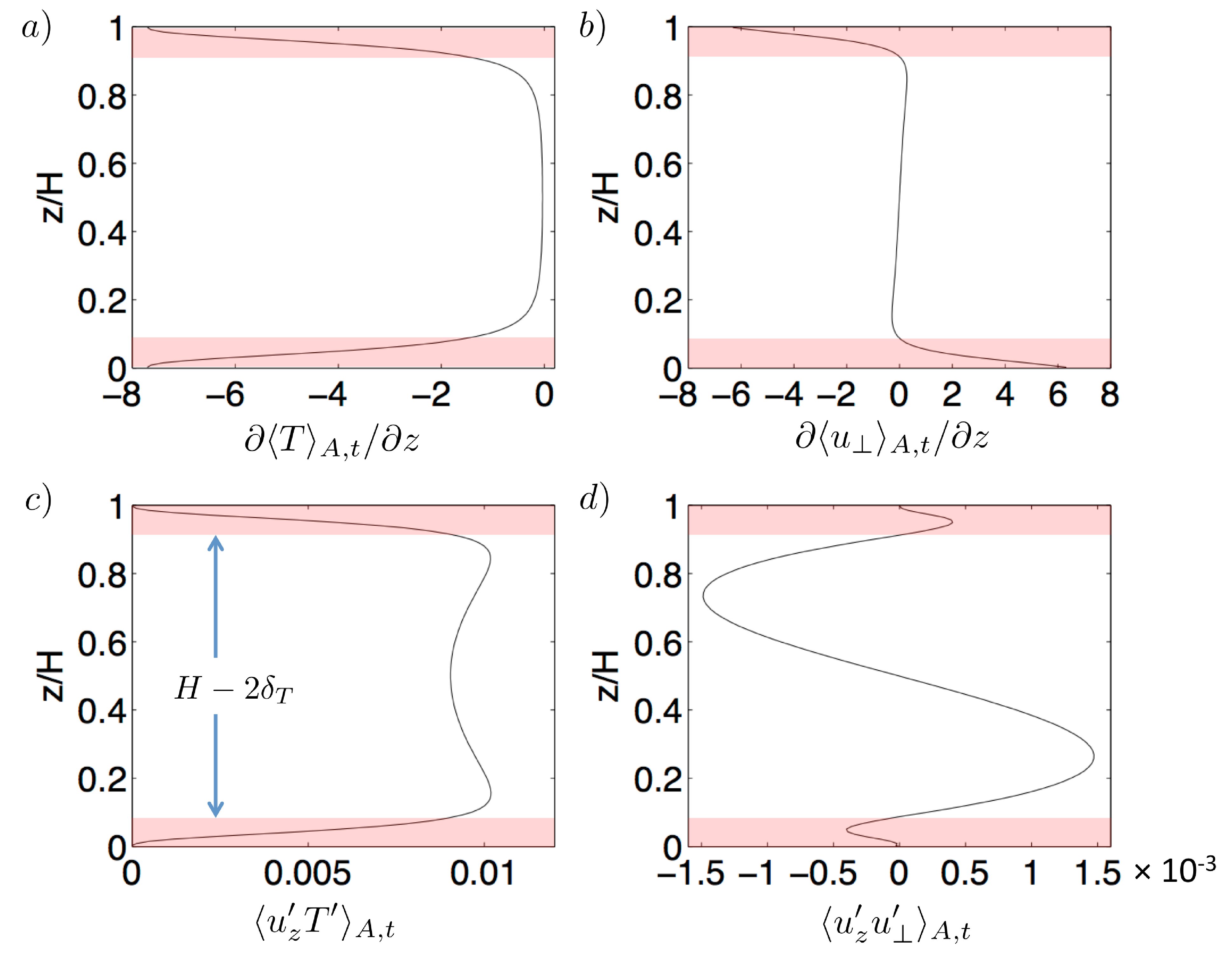}
\caption{(Color online) Vertical profiles of the plane and time averaged correlations and derivatives which are required to determine the turbulent viscosity and diffusivity.
Data displayed in the figure are obtained for $Ra=500\,000$ and $Pr=0.7$.}  
\label{fig4}
\end{center}
\end{figure}

In case of the turbulent diffusivity, we focus to the vertical transport of heat from the hot bottom plate to the cold top plate. The comparison of the three 
convective fluxes shows that the magnitude of the mean vertical flux is the largest. The turbulent diffusivity, $\kappa_{\ast}$,  can be obtained by the following bulk average 
\begin{equation}
\int_{\delta_T}^{1/2} \langle u^{\prime}_z T^{\prime} (z)\rangle_{A,t} \,dz=-\kappa_{\ast} \int_{\delta_T}^{1/2}\frac{\partial \langle T(z)\rangle_{A,t}}{\partial z} \,dz\,.
\label{kappa_turb}
\end{equation}  
Time-plane averages are denoted by $\langle\cdot\rangle_{A,t}$. Figure \ref{fig4} displays the resulting profiles which enter the determination of $\kappa_{\ast}$ via (\ref{kappa_turb}).
The double-headed arrow indicates the bulk region in the figure.

In case of the momentum transport the determination is less straightforward. We can expect that the horizontal turbulent mixing is also important. First, we proceed however
similar to the temperature field. We take the magnitude of horizontal velocity ${\bf u}_{\perp}(r,\phi,z,t)=u_r(r,\phi,z,t){\bf e}_r+u_{\phi}(r,\phi,z,t){\bf e}_{\phi}$. 
This field is decomposed again into a temporal mean and remaining fluctuations. The 
turbulent viscosity, $\nu_{\ast}$ is determined in a similar way as the turbulent diffusvity
\begin{equation}
\int_{\delta_T}^{1/2}  \langle u^{\prime}_z u_{\perp}^{\prime} (z)\rangle_{A,t} \, dz=-\nu_{\ast} \int_{\delta_T}^{1/2} \frac{\partial \langle u_{\perp}(z)\rangle_{A,t}}{\partial z}\, dz\,.
\label{nu_turb}
\end{equation}  
Here $u_{\perp}^{\prime}$ and $u_{\perp}$ denote magnitudes. The resulting profiles that enter (\ref{kappa_turb}) and (\ref{nu_turb})
are displayed in the right column of Fig. \ref{fig4}. In Tab. \ref{Tab1} we summarize the resulting turbulent Rayleigh and Prandtl numbers which result from this closure procedure. 
The turbulent Rayleigh numbers, $Ra_{\ast}$, are reduced for all three cases. $Ra_{\ast}$ gets consistently smaller with decreasing Prandtl number $Pr$ since 
the amplitude of the turbulent fluctuations increases. We obtain $Pr_{\ast} < Pr$ for all three cases. Their magnitudes vary between 0.2 and 0.4. The turbulent 
Prandtl number $Pr_{\ast}$ increases slightly with increasing $Pr$.
\begin{table}
\begin{center}
\begin{tabular}{ccccccccc}
\hline
Run & $Ra$ & $Pr$ &  $Ra_{\ast}$ & $Pr_{\ast}$ & $b_0$ & $b_1$ & $b_0^{\ast}$ & $b_1^{\ast}$\\
\hline
2 & $500\,000$  & 0.7  & $4\,500$  & 0.21 & 15 & 8 & 18 & 7 \\
5 & $500\,000$  & 10   & $39\,000$  & 0.38 & 20 & 1 & 18 & 1\\
\hline
\end{tabular}  
\end{center}
\caption{Betti numbers $b_0$ and $b_1$ for the original simulations at $Ra$ and $Pr$ as well as $b_0^{\ast}$ and $b_1^{\ast}$ for the corresponding runs at $Ra_{\ast}$ and $Pr_{\ast}$.  
The number of the runs corresponds with Tab. \ref{Tab1}. Temperature patterns at mid plane have been analyzed.}
\label{Tab2}
\end{table}

\begin{figure}
\begin{center}
\includegraphics[scale=0.18]{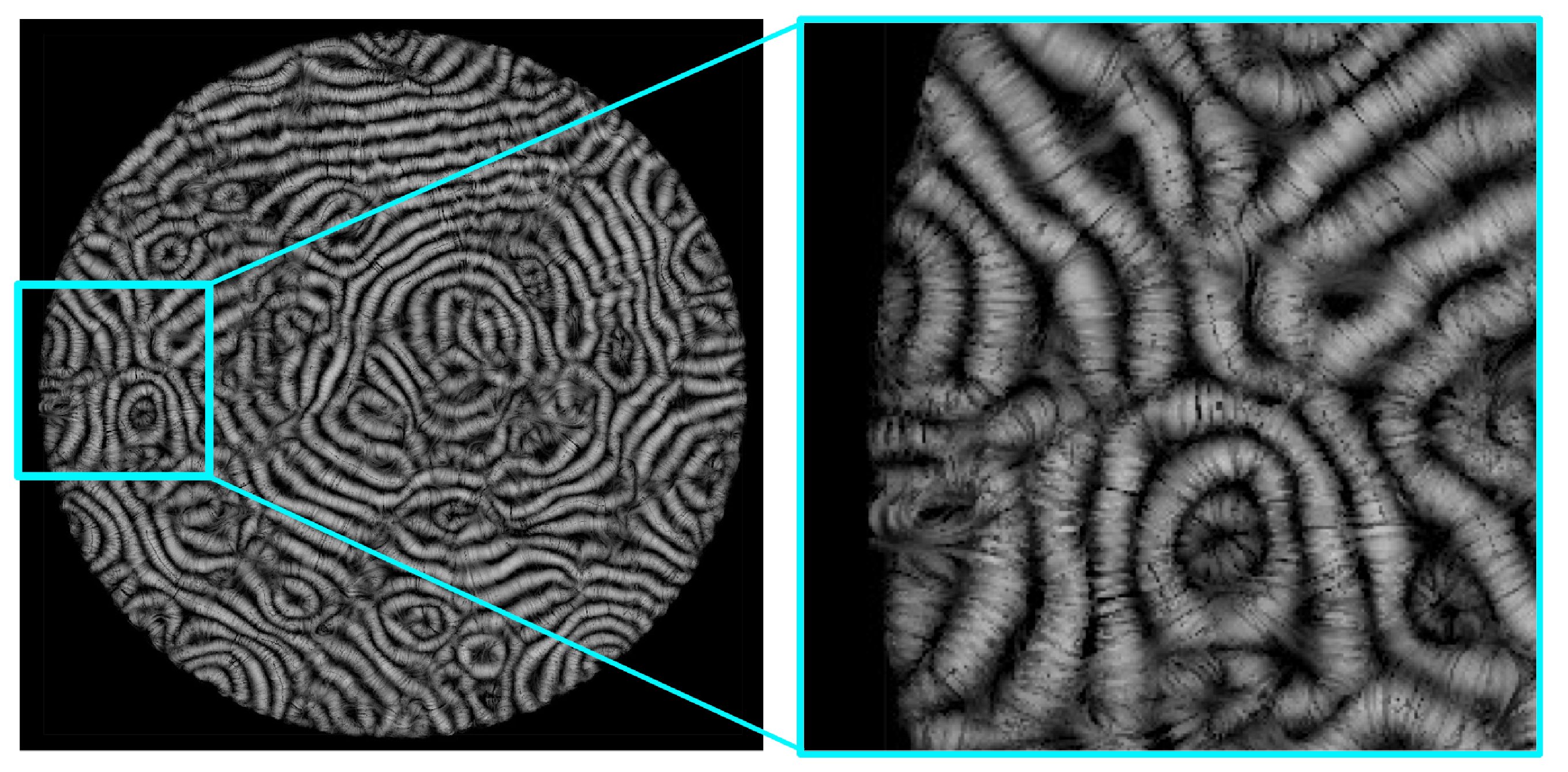}
\caption{(Color online) Streamline plots of the velocity field in a Rayleigh-B\'{e}nard convection cell as a view from the top. Left: plot averaged for 200 $T_f$.
Right: magnification of the same data. The DNS was conducted for $Ra=4\,500$, $Pr=0.2$ which corresponds with the turbulent viscosities
and diffusivities that correspond to the time-averaged data in Fig. \ref{fig1} (g,h).}  
\label{fig5}
\end{center}
\end{figure}

In case of runs 2 and 5, we then conducted a DNS with the same molecular viscosity and diffusivity as $\nu_{\ast}$ and $\kappa_{\ast}$, respectively.
The resulting streamline pattern for run 2 is displayed in Fig. \ref{fig5}. A time average over 200 $T_f$ was applied at $Pr=0.2$ and $Ra=4\,500$. 
The flow structure has to be compared now with the time-averaged one from Fig. \ref{fig1} (g,h) and indeed a reasonable visual agreement of both large-scale patterns is found.
We determined the Betti numbers $\{b_0, b_1\}$ from two-dimensional horizontal cuts
of the mean temperature at $z=1/2$ in both cases (\cite{Kurtuldu2011}). Betti numbers are $d$ positive integers to characterize a $d$-dimensional set topologically.
In detail, $b_0$ is the number of connected filaments which is obtained by digitizing a grayscale picture at a threshold, $b_1$ counts the number of enclosed holes in the pattern.  
We choose the temperature field in the mid plane.  A threshold temperature $T=0.5$ results in Betti number pairs which are listed in Tab. \ref{Tab2} for runs 2 and 5 as well as their 
corresponding runs at $Ra_{\ast}$ and $Pr_{\ast}$. Additionally, we estimated the average width of the rolls by counting the mean number of 
rolls that fit into the cell along different orientations. The values vary always around a width of $2H$, but are not exactly equal.

We also determined the turbulent viscosities from horizontal turbulent diffusion processes. It turns out that a simple adaption of the averaging 
procedure of Eqns. (\ref{kappa_turb}) and (\ref{nu_turb}) to a radial dependence is not successful. The roll patterns cause radially oscillating profiles which
result in strong cancellations for the averaged turbulent stresses and mean strain rates. If we omit the radial averaging and analyze
the local Boussinesq relation $\langle u^{\prime}_r u_j^{\prime} (z)\rangle_{\phi,H-2\delta_T,t}=-\nu_{\ast}(r) \partial \langle u_j(r)\rangle_{\phi,H-2\delta_T,t}/\partial r$
for $j=r,\phi,z$, we get indeed turbulent Prandtl numbers which are locally closer to one, but vary significantly with $r$. 

The resulting $Ra_{\ast}$ and $Pr_{\ast}$ are such that the DNS yield time-dependent patterns, in particular for run 5 with $Ra_{\ast}=39\,000$
and $Pr_{\ast}=0.4$. We therefore repeated this ``renormalization procedure'' in the weakly nonlinear regime and obtain $Ra_{\ast\ast}=3000$ and $Pr_{\ast\ast}=0.13$
for run 2 and  $Ra_{\ast\ast}=1800$ and $Pr_{\ast\ast}=0.17$ for run 5, respectively. Both runs end thus in the convection regime close to the onset.
  
To summarize this section, all routes of analysis will in general not lead to turbulent Prandtl numbers $Pr_{\ast}\approx 1$.  
A turbulent Prandtl number smaller than unity can be interpreted as follows: plume filaments of the temperature are coarser and diffuse faster than vortex filaments next to them. 
This circumstance could be connected to the fact that the width of rising and falling plumes is of the size of the thermal boundary layer thickness which is rather large for our $Ra$.
In contrast, vorticity is frequently generated on finer scales. Vortex filaments are for example generated by locally reversed flows next to rising plumes, a consequence of incompressibility.  An increase 
of the Rayleigh number to very large values could then increase the turbulent Prandtl number to one since the typical flow structures are getting finer and the boundary 
layers themselves are expected to become eventually fully turbulent.           

At this point, it should also be mentioned that the particular magnitudes of turbulent Prandtl numbers, $Pr_{\ast}$ are still an open problem.  
For example, \cite{Spiegel1971}, \cite{Kays1994}, or \cite{Groetzbach2011} discuss the dependence of $Pr_{\ast}$ on the distance from walls or on the 
original $Pr$. In case of homogeneous isotropic turbulence, \cite{Nakano1979} derived a value of $Pr_{\ast}=0.4$ from a spectral formulation 
based on the classical Kolmogorov turbulence theory.     
  
\subsection{Robustness of large-scale mean flow patterns to additional side wall forcing}
The sensitivity of SDC patterns to side wall effects and suppressed mean flows has been discussed in \cite{Bodenschatz1991} and \cite{Chiam2003}, respectively. This 
motivates us here, also in view to spontaneous symmetry breaking, to study their robustness with respect to an addition of a volume forcing to (\ref{nseq}).  The forcing is set up such that it sustains  
a steady Lamb--Oseen--type vortex $\hat{U}_i(r,z)$ with $i=\{r,z\}$ very close to the side walls of the cell at $(r_0=(\Gamma-1)/2, z_0=1/2)$. This vortex generates an
azimuthally symmetric mean flow at the side wall. The circulation, $\Omega$, and the radius of the vortex core, $r_L$, are chosen such that no-slip boundary conditions  can still be satisfied by setting this flow to
zero below a certain threshold. This clearly prohibits a stronger variation of the amplitudes and thus of the strength of the additional forcing.

Incompressibility of the full velocity field is sustained via the solution of the Poisson problem for the pressure in each time step.  
\begin{figure}
\begin{center}
\includegraphics[scale=0.18]{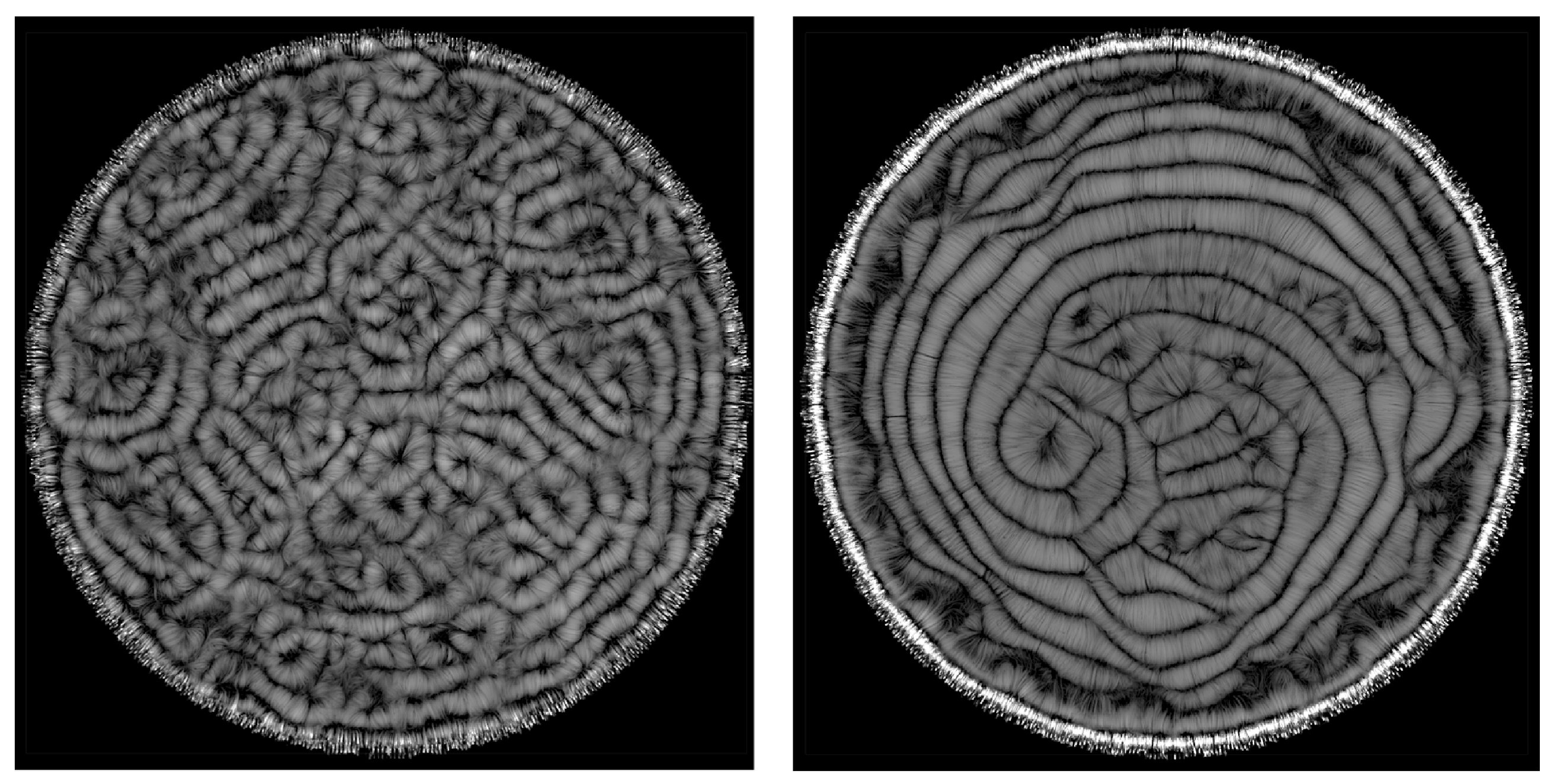}
\caption{Streamline plots of the velocity field in a Rayleigh-B\'{e}nard convection cell as a view from the top for $Ra=500\,000$ with the additional forcing 
$f_i$ (see also Eq. (\ref{nseq})). Left: $Pr=0.7$. Right: $Pr=10$. Both data sets have been averaged over $150 T_f$. We took $\Omega=1$ for the (non-dimensional)
circulation and $r_L=0.01$ for the (non-dimensional) radius of the vortex core.}  
\label{fig6}
\end{center}
\end{figure}
Figure \ref{fig6} shows the results for the mean flow pattern. In case of $Pr=0.7$, the toroidal roll is clearly visible right at the side wall. A second roll next to the side walls 
can be established by the additional forcing term. Towards the center of the convection cell the mean pattern remains however unchanged as can be seen by a comparison with panels (g,h) 
of Fig. \ref{fig1}. The turbulent fluctuations are large enough to re-establish the mean flow pattern. This is different for $Pr=10$. In comparison
to Fig. \ref{fig2}, the pattern has changed significantly. The toroidal roll pattern of the time averaged velocity is continued almost to the center of the cell. The reason 
for the stronger impact of the additional side wall forcing lies in the significantly lower level of turbulent fluctuations which we documented in Fig. \ref{fig3}. 

\section{Summary} 
We presented  three-dimensional DNS of thermal convection in the soft turbulence regime to study time-averaged velocity field patterns and their dependence on the Prandtl number
in very large aspect ratio convection cells. Our DNS demonstrate clearly that the SDC patterns, which 
are known from the weakly nonlinear regime, continue to exist in the turbulent regime. They remain thus dynamically relevant and do not simply disappear when convection
turns into the turbulent regime. The patterns are revealed when the turbulent fluctuations are removed by time averaging over intervals of the order of $10^2 T_f$, 
which is significantly smaller than the time scale over which the mean velocity and temperature patterns evolve. Our simulations allow us to calculate the 
turbulent viscosities and diffusivities as well as related turbulent Rayleigh and Prandtl numbers, $Ra_{\ast}$ and $Pr_{\ast}$. Their values fall indeed back into the range 
of the original SDC regime. The turbulent Prandtl numbers $Pr_{\ast}$ vary between 0.2 and 0.4 and increase with increasing $Pr$.  Our studies showed 
also that the mean patterns are robust to finite-amplitude perturbations once the turbulent fluctuations in the flow are sufficiently large, i.e., once $Pr$ 
at a given $Ra$ is sufficiently small. We demonstrated this by a side wall forcing that sustained an azimuthally symmetric vortex.  

Three future implications follow to our view: (i) it has to be investigated systematically if the mean flow patterns which are similar to SDC persist to even higher Rayleigh numbers
or if the mean flow structure is changed. This would require numerical studies at high Rayleigh numbers {\em and} large aspect ratios.
(ii) a more detailed analysis of the turbulent viscosities and diffusivities for larger Rayleigh numbers will provide useful input for technological and astrophysical 
applications in which the small-scale convective turbulence has to be modeled. This would however imply to explore systematically the tensorial nature of the turbulent viscosity
which we did not analyze in the present work. (iii) our results could also provide useful input to reduce the degrees of freedom systematically and to derive some effective
equations for the large-scale patterns, as done in other systems (\cite{Malecha2014}).  
  
\acknowledgements
This work is supported by the Deutsche Forschungsgemeinschaft.
Part of this work was completed while one of us (JS) stayed at the Institute of Pure and Applied Mathematics 
(IPAM) at the University of California Los Angeles. He thanks IPAM and the US National 
Science Foundation for financial support. Helpful comments by Janet Scheel and discussions 
with Jonathan Aurnou, Eberhard Bodenschatz, Friedrich Busse, Gregory Chini, and Keith Julien 
are acknowledged.

\end{document}